\documentclass[a4paper,fleqn,usenatbib]{mnras}
\usepackage{newtxtext,newtxmath}
\usepackage[T1]{fontenc}
\usepackage{ae,aecompl}
\usepackage{natbib,textcomp,placeins,epsfig,graphicx,color,amsmath,amssymb,upgreek,listings,epstopdf}
\usepackage{subfigure}

\newcommand{\logg}{$\log g$} 
\newcommand{\teff}{$T_{\mbox{\footnotesize eff}}$}
\newcommand{\al}{$\alpha$}

 \newcommand{\z}{$|z|$}
\newcommand{\Rgal}{$R_{\rm Gal}$}

\title[Spatial variations in the MAR]{Spatial variations in the Milky Way disc metallicity-age relation}

\author[D. K. Feuillet et al.]{Diane K. Feuillet,$^{1}$\thanks{email: feuillet@mpia.de} 
Neige Frankel,$^{1}$
Karin Lind,$^{1,2}$
\newauthor
Peter M. Frinchaboy,$^{3}$
D. A. Garc\'ia-Hern\'andez,$^{4,5}$ 
Richard R. Lane,$^{6,7}$
\newauthor
Christian Nitschelm,$^{8}$
Alexandre Roman-Lopes$^{9}$
\\
$^{1}$Max-Planck-Institut f\"ur Astronomie, K\"onigstuhl 17, D-69117 Heidelberg, Germany \\
$^{2}$Observational Astrophysics, Department of Physics and Astronomy, Uppsala University, Box 516, 571 20 Uppsala, Sweden \\
$^{3}$Department of Physics \& Astronomy, Texas Christian University, Fort Worth, TX 76129, USA \\
$^{4}$Instituto de Astrof\'isica de Canarias, E-38205 La Laguna, Tenerife, Spain \\
$^{5}$Universidad de La Laguna, Departmento de Astrof\'isica, E-38205 La Laguna, Tenerife, Spain \\
$^{6}$Instituto de Astrof\'isica, Pontificia Universidad Cat\'olica de Chile, Av. Vicuna Mackenna 4860, 782-0436 Macul, Santiago, Chile\\
$^{7}$Millennium Institute of Astrophysics, Av. Vicu\~na Mackenna 4860, 782-0436 Macul, Santiago, Chile\\
$^{8}$Centro de Astronom{\'i}a (CITEVA), Universidad de Antofagasta, Avenida Angamos 601, Antofagasta 1270300, Chile \\
$^{9}$Departamento de F\'isica, Facultad de Ciencias, Universidad de La Serena, Cisternas 1200, La Serena, Chile
}

\date{Accepted XXX. Received YYY; in original form ZZZ}

\pubyear{2019}

\begin{document}
\label{firstpage}
\pagerange{\pageref{firstpage}--\pageref{lastpage}}
\maketitle

\begin{abstract}

Stellar ages are a crucial component to studying the evolution of the Milky Way. Using Gaia DR2 distance estimates, it is now possible to estimate stellar ages for a larger volume of evolved stars through isochrone matching. This work presents [M/H]-age and [\al/M]-age relations derived for different spatial locations in the Milky Way disc. These relations are derived by hierarchically modelling the star formation history of stars within a given chemical abundance bin. For the first time, we directly observe that significant variation is apparent in the [M/H]-age relation as a function of both Galactocentric radius and distance from the disc mid-plane. The [M/H]-age relations support claims that radial migration has a significant effect in the plane of the disc. Using the [M/H] bin with the youngest mean age at each radial zone in the plane of the disc, the present-day metallicity gradient is measured to be $-0.059 \pm 0.010$ dex kpc$^{-1}$, in agreement with Cepheids and young field stars. We find a vertically flared distribution of young stars in the outer disc, confirming predictions of models and previous observations. The mean age of the [M/H]-[\al/M] distribution of the solar neighborhood suggests that the high-[M/H] stars are not an evolutionary extension of the low-\al\, sequence. Our observational results are important constraints to Galactic simulations and models of chemical evolution.

\end{abstract}


\section{Introduction}

The age-metallicity relation of the Milky Way disc has long been a focus of Galactic evolution studies \citep[e.g.][]{Twarog1980, Edvardsson1993, Casagrande2011} as any relation found would place tight constraints on models of Galactic chemical evolution (GCE). Simple `closed-box' models of galactic and stellar evolution dictate that over time the mean metallicity of a stellar population will increase as each generation of stars forms out of gas that has been enriched by previous generations. In reality, the absolute timeline of this metallicity enrichment is strongly dependent on the star formation history (SFH) as well as the amount and composition of gas that is injected into or ejected from the system (e.g. \citealt*{Chiappini1997}; \citealt{Dalcanton2007, Finlator2008}). In addition, the motions of stars are perturbed in such a way that over time they end up at a different mean distance to the Galactic center, become increasingly eccentric, and/or gain larger vertical oscillations \citep[e.g.][]{Schonrich2009}. 
Therefore an observational charactization of the full Milky Way disc age-metallicity relation provides a concrete result for potentially complex GCE models and cosmological simulations to replicate. 

The main difficultly in studying the Galactic age-metallicity relation has been the age determination of FGK field stars, which are the main targets of large spectroscopic surveys. The stellar metallicity and atmospheric parameters can be determined using spectroscopy, but the age cannot be directly measured. Traditionally, the most common method of determining stellar ages, especially for large samples, has been Bayesian isochrone matching based on the method described in \citet{Jorgensen2005}. This involves matching the observed properties of a star to models of stellar evolution to infer the most likely age. According to these models, after the first Gyr the observable surface properties of an FGK star change very little during its lifetime on the main sequence but change significantly as it begins to evolve beyond the core hydrogen burning phase. During the turn-off and subgiant phases,  the differences in the main observable properties of stars at a given metallicity and different masses (and therefore ages) are large compared to the spectroscopic measurement uncertainties in those parameters. This makes it possible to determine age with an uncertainty of $\sim 1$ Gyr through isochrone matching \citep[e.g.][among others]{Edvardsson1993, Casagrande2011, Bensby2014}. This is unfortunately not the case on the giant branch where stars of different masses and evolutionary stages can have very similar observable properties. 

Recent work has found that certain spectral features, such as Balmer lines \citep{Bergemann2016} or C/N ratios \citep{Masseron2015, Martig2016}, can trace the stellar mass, but this is also not a direct measurement of age. While data-driven and neutral network analyses do provide atmospheric parameter, abundance, and age estimates simultaneously from spectra, the age is mainly constrained by the empirically-derived relationship between the stellar mass and the element abundance information in the spectra \citep[e.g.][]{Ness2016b, Mackereth2019}. 

Asteroseismology can produce very precise ages for both dwarf and giant stars \citep[see][]{Gai2011, Chaplin2014} for regions of the Galaxy observed by {\it Kepler} \citep{Borucki2010, Koch2010}, {\it K2}  \citep{Howell2014}, or {\it CoRoT} \citep{Baglin2006a, Baglin2006b}. However, there currently exist many more high-resolution spectroscopic observations of stars outside these fields for which asteroseismology is out of reach. \citet[][hereafter F16]{Feuillet2016} show that is it possible to determine ages of giant stars to within $\sim0.18$ dex in log(age) through isochrone matching if precise distance measurements and high-resolution spectroscopy are available. By hierarchically modelling the star formation history (SFH), F16 and \citet[][hereafter F18]{Feuillet2018} derive [\al/M]-age and [M/H]-age relations for a sample of solar neighbourhood giants that are in good agreement with results from solar neighbourhood subgiants.

In recent years, the Milky Way disc age-metallicity relation has been examined using observations from large stellar surveys such as the Geneva-Copenhagen Survey \citep[GCS,][]{Casagrande2011}, the Gaia-ESO Survey \citep[GES,][]{Bergemann2014}, the Large sky Area Multi-Object Fiber Spectroscopic Telescope \citep[LAMOST,][]{Xiang2017}, Galactic Archaeology with HERMES \citep[GALAH,][]{Lin2018, Buder2019}, the Apache Point Observatory Galactic Evolution Experiment \citep[APOGEE,][F18]{Anders2017a, SilvaAguirre2018}, and the Multi-object APO Radial Velocity Exoplanet Large-area Survey \citep[MARVELS,][]{Grieves2018}. 
Interestingly, these studies have produced qualitatively similar age-metallicity relations although they have used different types of stars and different age determination methods. Generally, the local age-metallicity relation shows a large spread in metallicity at any given age, with a flat relation for young and intermediate age stars. The metal-poor stars are consistently older, but most studies find some metal-poor stars with intermediate ages and some old stars with solar metallicities. 

High-precision studies with smaller number statistics also find a large spread in metallicity at any given age, despite having small age uncertainties \citep[e.g.][]{Bensby2014, Haywood2013, SilvaAguirre2018, Nissen2018}. Such studies support the conclusion that the disc age-metallicity relation is intrinsically scattered and the spread is not solely an artefact of observational errors. In contrast, tight age correlations with [\al/Fe] or other individual elements have been found \citep[e.g.][F18]{Nissen2015, Bedell2018}. The age-metallicity relation is nevertheless interesting as the range of [M/H] values is larger than the range in [X/Fe] ratios. Additionally, it is important to confirm the spread in the age-metallicity relation throughout the Milky Way, which is difficult with smaller samples. 

The large range of ages covered by stars with a single metallicity and apparent lack of evolution in the age-metallicity relation has been the major focus of these studies; however, an equally interesting, and perhaps more diagnostic, feature is that the most metal-rich stars have intermediate ages. This is found in most studies of the solar neighbourhood, but is most striking in Figure 3 of F18, who examine the age distribution of mono-metallicity bins. The spread in age is large in each metallicity bin, but the mean age is nevertheless well defined with this technique and chemical evolution can be clearly seen from old metal-poor stars to younger solar metallicity stars. This trend is reversed at metallicities above solar, producing a turnover feature. The metal-rich stars are on average older than the solar metallicity stars. The hierarchical modelling technique of F16 and F18 relies on determining the mean age of a group of stars (in this case grouped by metallicity), therefore the trends derived are in fact metallicity-age relations (MAR).  

\citet{Minchev2013} use a GCE model in a full cosmological context, which includes radial migration, that reproduces these features of the observed local age-metallicity relations. Radial migration is the migration of stars inward or outward from their birth radius while maintaining circular, or near circular, orbits \citep[see][]{Wielen1996, Sellwood2002, Schonrich2009, Loebman2011}. This stellar migration is caused by the conservation of angular momentum during interactions with resonant features in the disc such as spiral arms. In the model of \citet{Minchev2013}, the stars currently at the solar Galactocentric radius were born at a range of radii, see their Figure 3. The metal-rich stars currently in the solar neighborhood were preferentially born in the inner Galaxy where star formation began earlier and proceeded more rapidly, resulting in metal-rich stars born earlier than was possible at larger Galactocentric radii.  GCE models that include radial migration can explain the large spread in metallicity at all ages because the birth age-metallicity relation is different at each Galactocentric radius, so as stars migrate inward or outward, they pollute the local age-metallicity relation of their new Galactocentric orbit  \citep[e.g.][]{Schonrich2009, Minchev2013, Kubryk2015}.

\citet{Mackereth2017} find a similar result observationally using APOGEE red giant branch stars and ages based on the mass-CN relation \citep{Martig2016}.  They show that mono-age, mono-[Fe/H] stars in the low-\al\, sequence have donut-like surface-mass density profiles. Stars younger than 3 Gyr are more tightly concentrated around the peak density, while the distribution of older stars is broader because they have radially migrated from their birth radius. The average metallicity of the young stars is a function of Galactocentric distance, with the metal-poor stars being concentrated in the outer Galaxy and the metal-rich stars being concentrated in the inner Galaxy.
 The radial metallicity gradient of the disc has been found by many previous studies using large spectroscopic surveys such as GCS \citep{Casagrande2011}, SEGUE \citep{Lee2011}, RAVE \citep{Boeche2013, Boeche2014}, GES \citep{Bergemann2014}, and APOGEE \citep{Hayden2014, Anders2014, Anders2017a}. Studies focusing on young stars to measure the gradient have used Cepheids \citep[e.g.][]{Genovali2014, Inno2019} and open clusters \citep[e.g.][]{Reddy2016, Donor2018}. The radial metallicity gradient supports the idea that the metal-rich stars with intermediate ages were likely born in the inner Galaxy.

While radial migration is predicted to have a significant effect in the plane of the disc, the process is less efficient for stars with larger vertical velocities, therefore it should have a smaller effect on stars at larger distances from the plane of the disc. This has been shown in dynamical models \citep[e.g.][]{Solway2012}, and simulations of galactic discs \citep[e.g.][]{Bird2013}. Observationally, \citet{Hayden2015} find that the shape of the metallicity distribution functions for stars through the disc are consistent with a metallicity gradient and a simple model of radial migration in the plane. Moving away from the plane, the metallicity distribution functions are uniform at all Galactocentric radii, perhaps reflecting a homogeneously mixed gas disc rather than a diffusive radial migration process.

\begin{figure*}
\centerline{
\includegraphics[clip, angle=0, trim=0cm 0cm 0cm 0cm, width=1.\textwidth]{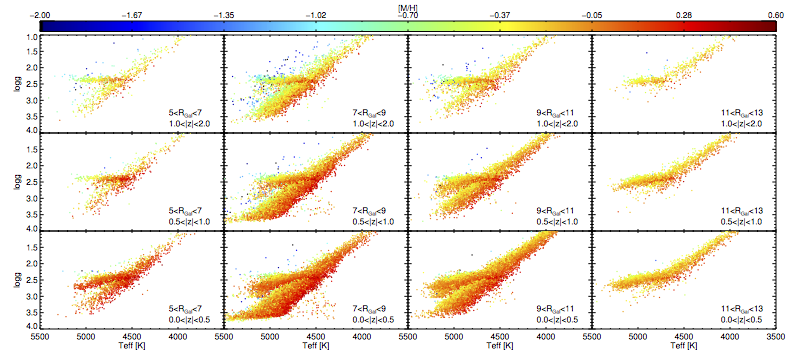} 
}
\caption{The spectroscopic Hertzsprung-Russell Diagram, or Kiel Diagram, for each of the 12 disc zones. The color indicates the metallicity ([M/H]). The \Rgal\, and \z\, bin are identified in the bottom right corner of each zone. }
\label{fig:HRD}
\end{figure*}

Until recently the samples of stars with age determinations have mostly been limited to the local disc or stars with chemically-based ages. But with the extensive parallax measurements provided by the second Data Release of the European Space Agency {\it Gaia} Mission \citep[DR2,][]{Gaia2018}, it is now possible to derive precise age trends for stars beyond the solar neighborhood. This work explores the spatial variations in the MAR of the  Milky Way disc. Using these variations we discuss the potential dependence of radial migration efficiency on scale height, the disc radial metallicity gradient, and the age evolution of the high-\al\, and low-\al\, sequences.

\section{Sample and Methods} \label{sec:method}

The sample presented is taken from the Data Release 14 \citep[DR14,][]{Abolfathi2018} of the Sloan Digital Sky Survey-IV \citep[SDSS-IV][]{Blanton2017} APOGEE \citep[][]{Majewski2017} for which {\it Gaia} Data Release 2 \citep[GDR2,][]{Gaia2018} parallax measurements are available. APOGEE is an $H$-band spectroscopic survey of Milky Way stars using the 300-fibre fed, high-resolution ($R \sim$ 22,500), APOGEE spectrograph \citep{Wilson2019} on the Sloan Foundation 2.5-m telescope \citep{Gunn2006}. APOGEE data are reduced using a standard pipeline \citet{Nidever2015} and analyzed by the APOGEE Stellar Parameters and Chemical Abundances Pipeline \citep[ASPCAP,][]{GarciaPerez2016}. In DR14, ASPCAP uses a specially computed library of 1D plane parallel stellar atmospheric models assuming local thermodynamic equilibrium \citep{Zamora2015}, which is calculated using a custom built line-list \citep{Shetrone2015}, to simultaneously determine effective temperature (\teff), surface gravity (\logg), metallicity ([M/H]), [\al/M], [C/M], and [N/M] through $\chi^2$-minimization. Calibration and quality analysis of DR14 are described in \citet{Holtzman2018} and \citet{Jonsson2018}. The targeting scheme for APOGEE is described in \citet{Zasowski2013} and \citet{Zasowski2017}. 

The following selection and quality criteria have been applied in this work:
\begin{itemize}
\item $\sigma_{\pi}/\pi < 0.2$
\item $(J-K)_0 \ge 0.5$
\item 1.0 < \logg < 3.8
\item 3500 < \teff < 5500
\item SNR > 80.
\item No \verb STAR_BAD  or \verb VSINI_WARN  ASPCAP flag set
 \item No \verb BAD_PIXELS  or \verb VERY_BRIGHT_NEIGHBOR  star flag set
\item No known or candidate cluster members
\end{itemize}
 The final sample contains 81,400 stars. The results presented have removed known cluster members and are limited to stars with radial Galactocentric distances (\Rgal) between 5 and 13 kpc and a distance from the plane (\z) of less than 2 kpc, resulting in 77,562 stars. To explore the spatial variation in the derived age trends, the sample is divided into four \Rgal\, bins (5-7, 7-9, 9-11, and 11-13 kpc) and three \z\, bins (0.0-0.5, 0.5-1.0, and 1.0-2.0 kpc), resulting in 12 disc zones. With increasing \z, the zones become dominated by stars with positive $z$ height, but we find the MARs are symmetric in $z$. We therefore use \z\, to increase the signal in all zones. Figure \ref{fig:HRD} shows the \logg-\teff\, diagram for each of the 12 zones with the [M/H] indicated by the color. The full red giant branch is sampled in all zones except \Rgal\, 11-13 kpc where the high \logg\, stars are too faint for the {\it Gaia} parallax restriction.

\subsection{Age determination}

The age trends presented here were determined using the hierarchical modelling method described in F16 and F18. This method constrains the parameters of a model SFH by combining the age likelihood functions produced from Bayesian isochrone matching for a group of stars. The parameters used in the Bayesian isochrone matching are \teff, \logg, overall metallicity, and absolute $K$ magnitude. The overall metallicity is calculated from the APOGEE [M/H] and [\al/M] using the prescription of \citet{Salaris1993}. The stellar atmospheric parameters (\teff, \logg, [M/H], and [\al/M]) are calibrated APOGEE DR14 values. A full description of the calibrations applied to DR14 is available in \citet{Holtzman2018} and \citet{Jonsson2018} provides a thorough comparison of APOGEE abundances to independent analyses. The absolute $K$ magnitudes are calculated using the $K$-band magnitudes from the Two Micron All Sky Survey \citep{Skrutskie2006}, a distance, and an extinction. The distance value is taken from \citet{Bailer-Jones2018}. The extinction is taken from the APOGEE targeting information using the \verb AK_TARG \, parameter \citep[see][]{Zasowski2013} which uses the RJCE method \citep{Majewski2011}. If the \verb AK_TARG \, parameter is not available, then the WISE $K$-band extinction is used from the \verb AK_WISE \, parameter. In this work, PARSEC isochrones \citep{Bressan2012} are used with a lognormal Chabrier initial mass function \citep{Chabrier2001}.  We note that PARSEC isochrones do not include atomic diffusion, which can affect the surface metallicity, but this process is thought to have a minimal effect on giant stars \citep{Dotter2017}.

The model SFH used is a Gaussian function plus a uniform component allowing for outliers within a group of stars assuming an outlier fraction of 7.5~\% as in F16. For this work, the stars are grouped by abundance, specifically [M/H] and [\al/M]. The widths of the abundance bins are determined by the mean uncertainty in the abundance measurements. If fewer than 15 stars lie within the bin then the width is increased until it contains 15 stars. The result of the hierarchical modelling is a mean age and age dispersion for the stars in a given abundance bin and can be applied to individual abundances independently. Due to the method of binning stars in abundance starting with the stars with the lowest abundances, the lowest abundance bin usually contains stars that cover a larger abundance range and are typically outliers from the main abundance distribution. 

\begin{figure}
\centerline{
\includegraphics[clip, angle=0, trim=1cm 0cm 1cm 1cm, width=0.48\textwidth]{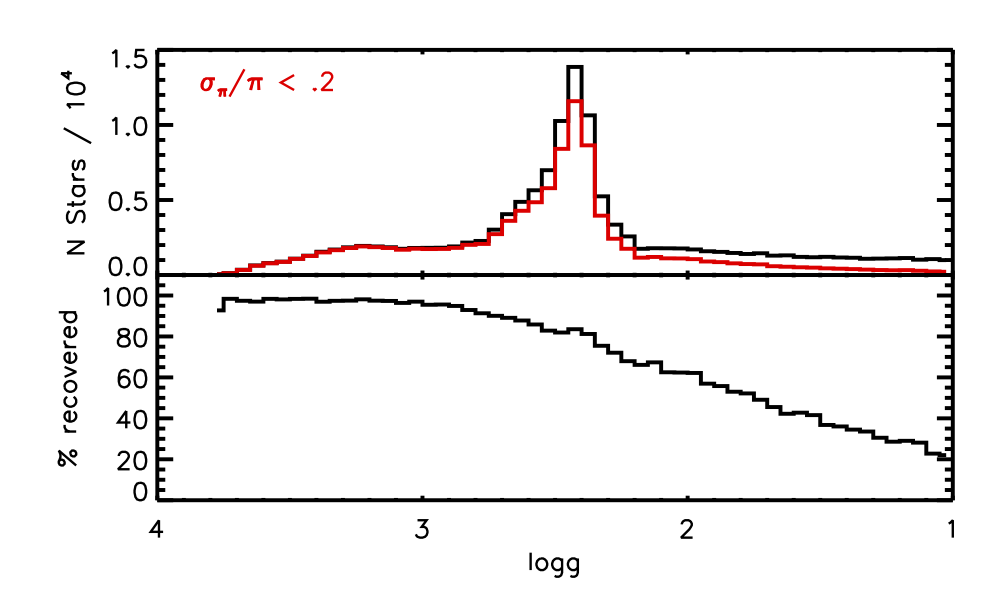}
}
\caption{The fraction of stars included in the sample as a function of \logg\, as compared to the full sample without any restriction on parallax uncertainty.}
\label{fig:loggbias}
\end{figure}

\begin{figure*}
\centerline{
\includegraphics[clip, angle=0, trim=0cm 0cm 0cm 0cm, width=1.0\textwidth]{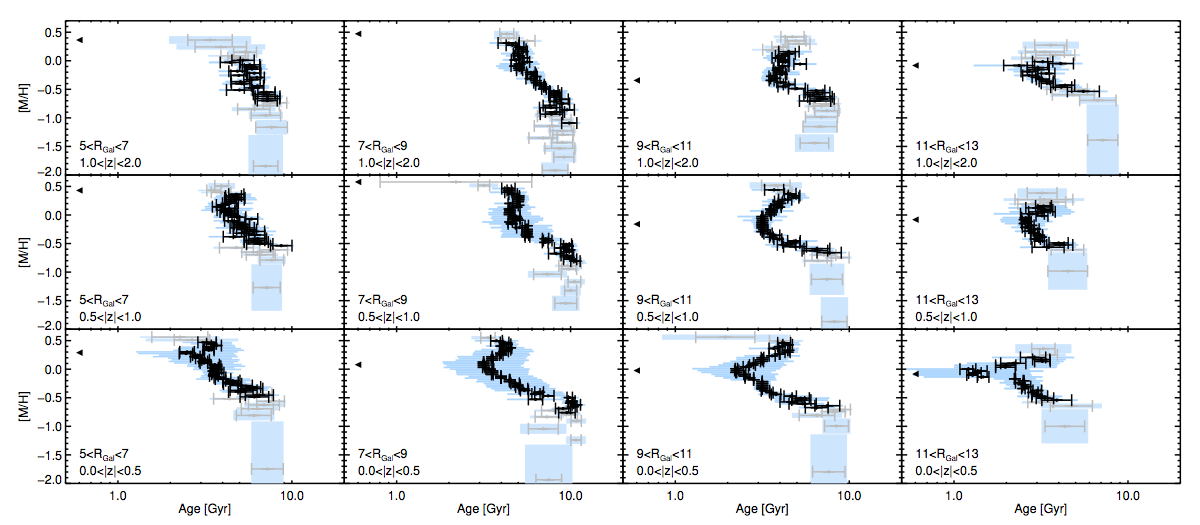}
}
\caption{The metallicity-age relation in each of the 12 disc zones. The black points indicate the mean age of each [M/H] bin and the error bar shows the uncertainty. The blue shaded region designates the age dispersion of the [M/H] bin. Bins with only 15 stars are lighter in color than the other bins. The \Rgal\, and \z\, bin is identified in the bottom left corner of each zone. The metallicity of the youngest bin in each zone is shown by the left-facing triangle.}
\label{fig:AMR}
\end{figure*}

We apply a correction for the bias in \logg\, imposed by the 20~\% cut in parallax uncertainty. The low \logg\, stars are intrinsically bright and are therefore preferentially farther away in a survey selecting targets from a limited magnitude range. At a given apparent magnitude, more distant stars have larger parallax uncertainties than closer stars. These effects lead to a bias against low \logg\, stars in this sample. The \logg\, bias in this sample is illustrated in Figure \ref{fig:loggbias}. The top panel shows the \logg\, distribution of the sample with no parallax uncertainty cut in black, and imposing a maximum of 20~\% in parallax uncertainty in red. The bottom panel shows the fraction of stars recovered after imposing the 20~\% cut as a function of \logg. While the parallax uncertainty limit has almost no effect on the high \logg\, stars, up to 80~\% of the low \logg\, stars are removed by this selection. This selection bias is accounted for during the normalization of the SFH model, see $N(a)$ in section 4.6 of F16. 

 A selection function that accounts for the APOGEE-1 color selection and the cuts we imposed on \teff\, and \logg\, is also applied during the hierarchical modelling. This does not account for the full APOGEE - {\it Gaia} crossmatch selection function, which is more complex and beyond the scope of this paper. A full treatment of the survey selection functions is planned for future comparisons with Galactic models where accounting for possible biases will be crucial. Here we discuss generally the biases expected due to the APOGEE-1 and APOGEE-2 targeting strategies \citep{Zasowski2013, Zasowski2017}. APOGEE-1 used a single color selection for the disc and bulge fields, $(J-K)_0 \ge 0.5$, but APOGEE-2 applies a dual color selection, $0.5 \le (J-K)_0 \le 0.8$ and $(J-K)_0 \ge 0.8$. In both APOGEE-1 and APOGEE-2 the halo fields use a color selection of $(J-K)_0 \ge 0.3$, but our full sample color cut in this work is also applied to the halo fields, so no age bias is expected. Possible age biases caused by the APOGEE-2 dual color selection are expected to be up to $\sim0.1$ dex older in the disc and bulge fields.

The most distant zones are biased towards luminous, upper giants branch stars, as seen in Figure \ref{fig:HRD}, due to the parallax uncertainty cut. This luminosity bias is expected to cause a bias towards younger ages, up to $\sim0.15$ dex, in the most distant zones. Using the solar neighborhood sample, we find that there is no bias in [M/H] or [\al/M] due to the lack of lower giant branch stars. We therefore determine that the shape of the outer zone abundance-age relations should be unaffected. 
This luminosity age bias is opposite to the effects expected from the color bias in the disc and bulge fields.
In Section \ref{sec:results} we discuss how these biases may affect our interpretation of results.


\section{Age trends} \label{sec:results}

\subsection{Metallicity-Age Relation} \label{sec:amr}

The age-metallicity relation in the solar neighborhood has been observed to have stars with a large spread in metallicity at any given age \citep[e.g.][]{Edvardsson1993, Casagrande2011, Bensby2014}. Most recent studies find that the most metal-poor stars have the oldest ages, the most metal-rich stars have intermediate ages, and the youngest stars have solar metallicities \citep[see][F18]{Casagrande2011}. These observed deviations from the narrow and monotonic age-metallicity relation predicted by simple models of chemical evolution have been tentatively attributed to radial migration of stars in the Galactic disc (F18). With the sample presented here it is possible to search for detailed spatial variations in the MAR through the Galactic disc. Again, we note that the analysis presented determines a mean age for stars binned in [M/H]. Comparisons are only made to unbinned literature age-metallicity relation results. 

\begin{figure*}
\centerline{
\includegraphics[clip, angle=0, trim=0cm 0cm 0cm 0cm, width=1.0\textwidth]{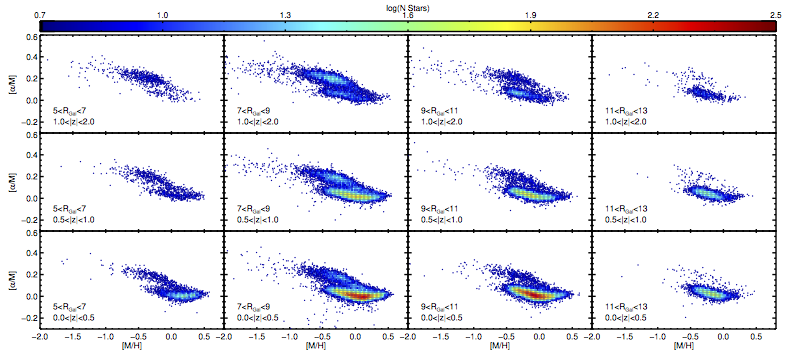}
}
\caption{The [\al/M] vs [M/H] distributions for each of the 12 disc zones. Color indicates the logarithmic number density. The \Rgal\, and \z\, bin are identified in the bottom left corner of each zone.}
\label{fig:xm_al}
\end{figure*}

Figure \ref{fig:AMR} shows the hierarchically modelled MAR for 12 disc zones. 
The points mark the mean age derived for each abundance bin and the error bars indicate the uncertainty in the mean age. The blue shaded regions show the age dispersion and abundance width of each bin. Bins containing only 15 stars are given a lighter color. From this figure is it clear that significant spatial variations exist in the Galactic MAR.
To put these MARs in context with the [\al/M] vs [M/H] abundance distributions, Figure \ref{fig:xm_al} shows the [\al/M] vs [M/H] distribution for each of the 12 zones. These distributions show that the high-\al\, sequence is dominant at high \z\, and has a shorter scale length than the low-\al\, sequence, which dominates in the plane of the disc and is strongly present out to larger \Rgal. The relative spatial distributions of the high- and low-\al\, sequences shown here are in agreement with APOGEE DR12 results from \citet{Hayden2015} and the APOGEE DR14 [Fe/Mg] vs [Mg/H] distributions shown in \citet{Weinberg2019}.

Starting in the solar neighborhood, the clear turnover behavior of the local MAR noted by F18 is recovered in this larger sample. This sample contains 15,180 stars in the $7 <$ \Rgal $ < 9$, $0 < \z < 0.5$ zone, while F18 presented 721 stars within 400 pc of the Sun. The increase in mean age with decreasing metallicity below [M/H] $\sim 0$ extends to [M/H] $\sim -1$. The youngest stars in this zone have a metallicity of -0.1 to 0.2, consistent with the local interstellar medium \citep{Nieva2012}, and most were likely born within the zone. The Sun is older than mean age of solar [M/H] stars in the solar neighborhood. Assuming an age of $4.66$ Gyr \citep{Dziembowski1999} and [M/H] of 0, Figure \ref{fig:AMR} suggests that the Sun was likely born inward of its current \Rgal\, between 5 and 7 kpc, or even closer to the Galactic centre. This is consistent with previous estimates \citep{Wielen1996, Minchev2013, Minchev2018}.
The metal-rich stars, which are older than the solar metallicity stars and more evolved chemically, probably radially migrated from elsewhere in the disc.   The presence of intermediate-age, metal-rich stars in the solar neighborhood is a generic result of GCE models that include radial migration. In such models, these metal-rich stars form in the inner disc, see figure 5 of \citet{Minchev2014}. In \citet{Kubryk2015}, this results in a MAR with a similar turnover feature as the present work when considering the mean age of all stars, not just those formed {\it in situ}, see their figure 12.

One of the most striking features of Figure \ref{fig:AMR} is the presence of the mean age turnover of high metallicity stars in all of the $0 < \z < 0.5$ zones (bottom row of Figure \ref{fig:AMR}). This suggests that the radial migration of stars in the disc could be a significant process \citep[e.g.][]{Minchev2014, Frankel2018, Weinberg2019}. In this larger sample, there is also a secondary turnover feature around [M/H] $\sim0.4$ dex. The cause of this feature is unknown, but could reflect the intrinsic age-metallicity relation of the inner disc if these highest metallicities are significantly dominated by stars migrated from a similar birth \Rgal. The presence of a second turnover feature is suggested by figure 15 of F18, which uses a mixture of analytic chemical evolution models representing multiple zones of chemical evolution to simulate a population of stars born over a range of \Rgal.  This second turnover is also seen in figure 12 of \citet{Kubryk2015} and is caused by stars above [Fe/H] = 0.4 coming from the inner $2-3$ kpc.

Interestingly, the metallicity of the primary turnover changes as a function of \Rgal. The metallicity of the youngest stars in each zone is shown by the left-facing triangles in Figure \ref{fig:AMR}. This is a confirmation of the disc radial metallicity gradient predicted by simulations and observed using Cepheids \citep[e.g.][]{Genovali2014, Inno2019}, open clusters \citep[e.g.][]{Reddy2016, Donor2018}, and field stars \citep[e.g.][]{Boeche2013, Bergemann2014, Hayden2014}. The confirmation of the metallicity gradient supports the hypothesis that the most metal-rich stars in the plane of the disc likely came from the inner Galaxy, \Rgal $< 5$ kpc, and the assumption used in many GCE models that star formation began earlier and proceeded more rapidly in the inner Galaxy.

Using the youngest mean age and mean \Rgal\, of these four radial zones in the plane of the disc, we estimate the metallicity gradient to be $-0.061 \pm 0.015$ dex kpc$^{-1}$. If we use \Rgal\, bins of 1 kpc instead of 2 kpc, the measured metallicity gradient is $-0.059 \pm 0.010$ dex kpc$^{-1}$. These two measurements are consistent, but higher precision is reached with a finer \Rgal\, binning. Our gradient measurement is consistent with measurements from Cepheids by \citet{Genovali2014} and \citet{Lemasle2007}, recent open cluster measurements by \citet{Donor2018},  APOGEE measurements of young field giants with ages determined using [C/N] abundances \citep{Hasselquist2018} and asteroseismology \citep{Anders2014, Anders2017a}, young field dwarfs observed by GES \citep{Bergemann2014} and RAVE \citep{Boeche2013}, and the present-day gradient reported by \citet{Minchev2018} using young field subgiant and turn-off stars. However, previous studies using open clusters and giant stars of all ages measure $\sim 0.08 - 0.1$ dex kpc$^{-1}$ \citep[e.g.][]{Frinchaboy2013, Hayden2014, Jacobson2016}. This difference could come from increased sample sizes, improved distance measurements, and an emphasis on using young stars in the other studies.

\begin{figure*}
\centerline{
\includegraphics[clip, angle=0, trim=0cm 0cm 0cm 0cm, width=0.50 \textwidth]{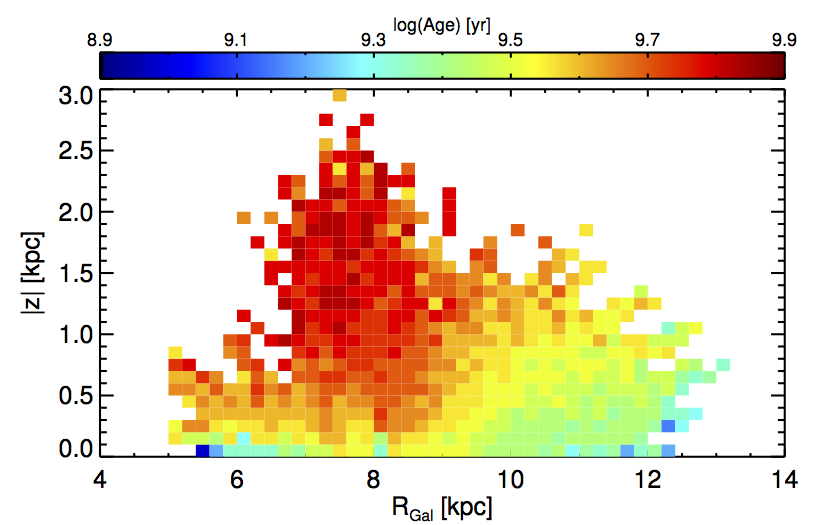}
\includegraphics[clip, angle=0, trim=0cm 0cm 0cm 0cm, width=0.50 \textwidth]{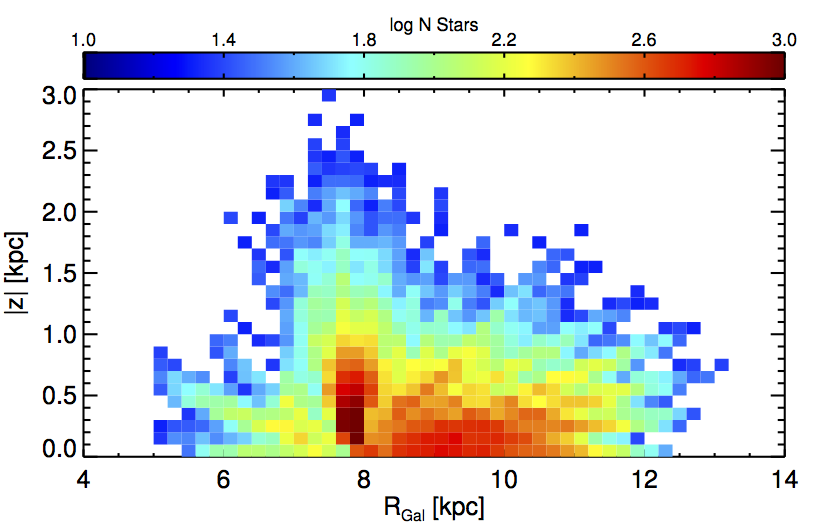}
}
\caption{The \z\, vs \Rgal\, distribution colored by the hierarchically modelled mean age (left) and logarithmic number density (right). The sample is binned by 200 pc in \Rgal\, and 100 pc in \z\,. Only bins that contain at least 15 stars are shown. The young stars have a flared distribution in the outer disc.}
\label{fig:flare}
\end{figure*}

In the $7 <$ \Rgal $< 9$ kpc bin, at larger distances from the disc plane, the turnover softens and almost flattens in the $1 < \z < 2$ kpc bin where the mean age does not vary much in different metallicity bins. The softening in the turnover with distance from the disc plane could suggest that stellar migration is much more efficient in the plane of the disc and has only a small effect at large \z. As seen in Figure \ref{fig:xm_al}, there are also fewer metal-rich stars in the highest \z\, zone, again consistent with the picture that the higher \z\, zones do not contain many radially migrated stars. 
However, the change in the turnover feature with increasing \z, could also be caused by a lack of young stars at high \z. The $7 <$ \Rgal $< 9$, $1 < \z < 2$ bin is dominated by the high-\al\, sequence, which has been found to be uniformly old \citep[see Figure \ref{fig:MAA} and ][]{Xiang2017}. 

With increasing \z, the youngest stars are no longer present in the inner disc, \Rgal $< 9$, and the zones are dominated by older, high-\al\, stars. At these \Rgal\, the young stars are born kinematically cold and close to the plane of the disc. With time their vertical velocities may increase, allowing them to spend time at larger \z\, where the older, high-\al\, sequence dominates. The $5 <$ \Rgal $ < 7$, $1 < \z < 2$ zone is almost entirely dominated by the high-\al\, sequence in this sample, and shows very little change in mean age as a function of [M/H]. This suggests that the high-\al\, sequence is either well mixed or formed all at once.

In the disc plane, the mean age of the metallicity bin with the youngest stars decreases as a function of \Rgal\, from 2-3 Gyr in the inner two zones to 1-2 Gyr in the outer two zones. There is an increased dominance of young stars in the outer disc at all \z\, heights. This dominance of young stars in the outer Galaxy has been observed using APOGEE \citep{Ness2016b} and LAMOST \citep{Xiang2017}. It is likely related to the decreased relative contribution in the outer disc of the high-\al\, sequence, which has been found to be generally older than the low-\al\, sequence. 
 While we expect some selection effects are present, discussed in Section \ref{sec:method}, we estimate that the net age bias is quite small. The low \z\, zones could be biased towards older ages, which would only amplify the trends observed with \z. 
The $11 <$ \Rgal $ < 13$ zones would be most significantly affected by the luminosity bias towards younger ages, but we note that the presence of young stars at larger \z\, is clearly seen in the $9 <$ \Rgal $ < 11$ zones as well.

Overall, the spatial variations in the MAR are in excellent agreement with the age trends found by \citet{Hasselquist2018} inferred using APOGEE [C/N] abundance ratios, see their figure 4. It has been shown that the [C/N] of giant stars correlates with the mass, and therefore the age, due to internal mixing of CN-cycle processed material from the core \citep{Masseron2015, Martig2016}. This relation was confirmed using APOGEE DR14 data by \citet{Hasselquist2018} and F18. \citet{Hasselquist2018} find a turnover in the [C/N] at high [Fe/H], corresponding to a turnover in age, in the plane of the disc that weakens with \z. They also find evidence of fewer old stars in the outer disc.

\subsection{Flared young disc} \label{sec:flare}

In the present sample, there is an increased fraction of young stars at larger distances from the mid-plane in the outer Galaxy compared to the inner Galaxy. In Figure \ref{fig:AMR}, the outer disc shows young stars present at larger \z\, zones than in the inner disc, reflecting the flared age distribution of the outer disc observed using stellar ages by \citet{Ness2016b} and \citet{Xiang2017}, and predicted by simulation \citep[e.g.][]{Minchev2014, Minchev2015, Rahimi2014}. \citet{Bensby2011} also note an absence of \al-rich stars in the outer disc, consistent with both Figure \ref{fig:xm_al} and an increased fraction of young stars in the outer disc.

To better illustrate the flared distribution of the young stars, we hierarchically model the mean age of the sample as a function of \Rgal\, and \z. Figure \ref{fig:flare} shows the \z\, vs \Rgal\, distribution of the sample colored by the mean age of each bin in the left panel and by the number of stars in each bin in the right panel. The distribution is binned by 200 pc in \Rgal\, and 100 pc in \z. As in Figure \ref{fig:AMR}, bins are required to contain at least 15 stars. In this case the bin size is not increased in order to contain 15 stars, the bin is simply not shown. The mean age of the outer disc is younger than the inner disc. In particular, inwards of 9 kpc, the young stars dominate at \z\, less than 300 pc. Beyond \Rgal\, of 9 kpc, the young stars dominate at larger \z, reaching 1 kpc at \Rgal\, of 12 kpc.  This is consistent with inside-out disc formation \citep[e.g.][]{Bird2013}, but inspection of the density profiles of mono-age populations is needed to confirm the true spatial distribution of young stars.

\begin{figure*}
\centerline{
\includegraphics[clip, angle=0, trim=0cm 0cm 0cm 0cm, width=1.0\textwidth]{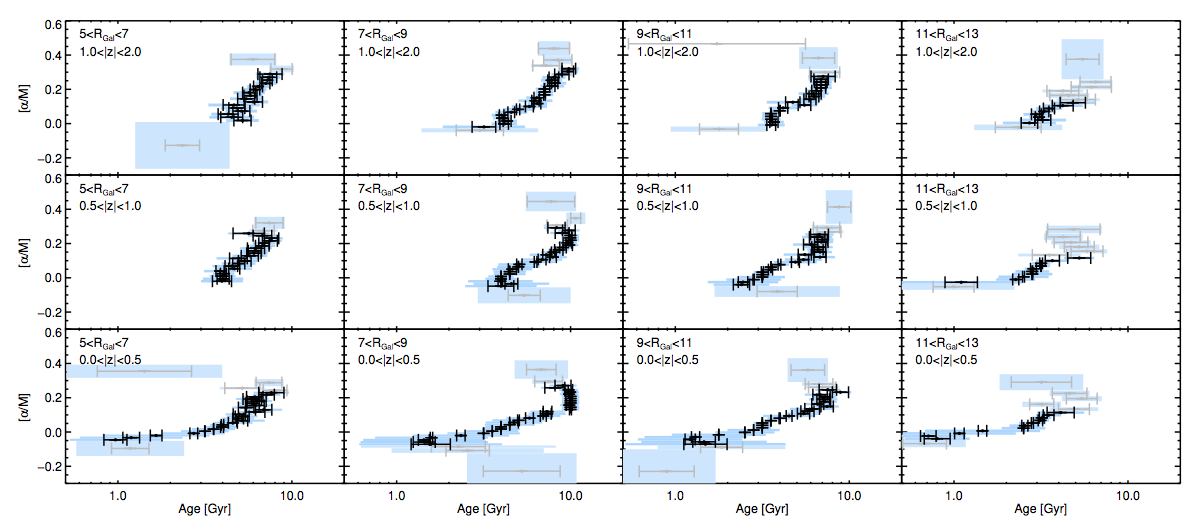}
}
\caption{Same as Figure \ref{fig:AMR} but binned in [\al/M].}
\label{fig:AAR}
\end{figure*}

 The luminosity bias in the outer disc is expected to be counteracted by the color bias in the disc fields. At higher \z\, the density of stars is low and not strongly represented in this figure. The flared behavior is also clear even at 9-10 kpc.
The flared age structure is suspiciously centered near the solar \Rgal\, (8 kpc) and could be influenced by some selection effects. However, the overall shape is similar to \citet[see their figure 23]{Xiang2017} using LAMOST data with different selection criteria. We preformed the same analysis on a sample of only APOGEE-1 stars (for which the single color selection has been taken into account) and the resulting age distribution is the same as Figure \ref{fig:flare}.


\subsection{[\al/M]-Age Relation} \label{sec:aar}

Figure \ref{fig:AAR} is the same as Figure \ref{fig:AMR}, but binned in [\al/M] abundance instead on [M/H]. The solar neighborhood [\al/M]-age relation (AAR) in this sample is in good agreement with the AAR presented in F18. The mean age increases rapidly with increasing [\al/M] at low abundances, but at [\al/M] above 0.15 dex, the mean age is approximately constant. Other \Rgal\, zones in the disc plane have a smoother transition in the AAR from low-\al\, to high-\al\, stars bins, but most zones show a transition around [\al/M] $\sim0.1$ dex, above which the high-\al\, sequence dominates. This transition in AAR suggests that the high-\al\, sequence and the low-\al\, sequence had very different chemical enrichment histories.  The large age evolution at low [\al/M] and smaller age evolution at high [\al/M] is consistent with most models of GCE \citep[e.g.][]{Kubryk2015} and observations of the local Galaxy by \citet{SilvaAguirre2018} using asteroseismology in the {\it Kepler} field as well as \citet{Haywood2013} using solar neighbourhood subgiants. 

The turnover towards older ages in the lowest abundance bins of the $7 <$ \Rgal $ < 9$, $0 < \z < 0.5$ zone is reminiscent of the age trends with Si, S, and Ca noted by F18. As in F18, the stars populating these abundance bins are mainly outliers in the [\al/M] vs [M/H] distribution and are not strongly present in other spatial zones. It is possible that these stars have been accreted from a merger, such as Sagittarius, and do not belong to the disc population. The highest [\al/M] bins in this zone have mean ages that are younger than the bin at 0.2 dex. The cause of this is unknown, but these bins contain fewer stars and have large mean age uncertainties. 

The AAR is very similar in all the spatial zones; the main differences arise due to the presence of young stars. This is apparent in the $7 <$ \Rgal $< 9$ zone. At farther distances from the midplane the young stars are no longer present, as noted in the MAR. This results in older mean ages for the low [\al/M] bins, making the full relation appear steeper. The AAR for the [\al/M] bins above $\sim 0.1$ is similar in all zones suggesting that the high-\al\, sequence is fairly uniform across the disc. In the outer disc, the young stars are present at larger \z\, than in the inner disc, again lending evidence to the flared distribution of young stars in the outer disc. The outer disc reaches mean ages of only 4-5 Gyr. As discussed above, the high-\al\, sequence is not strongly present in the outer disc and it is unlikely that star formation rates were high at these \Rgal\, in the early Milky Way.  As in Figure \ref{fig:AMR}, the outer disc zones are most likely to be biased by selection effects, but the $9 <$ \Rgal $ < 11$ zones already have an increased presence of young stars.


\begin{figure*}
\centerline{
\includegraphics[clip, angle=0, trim=0cm 0cm 0cm 0cm, width=0.48 \textwidth]{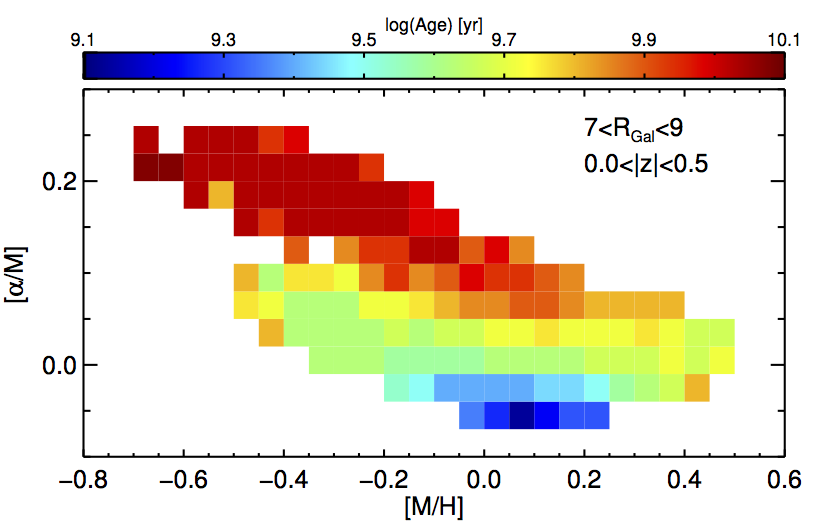}
\includegraphics[clip, angle=0, trim=0cm 0cm 0cm 0cm, width=0.48 \textwidth]{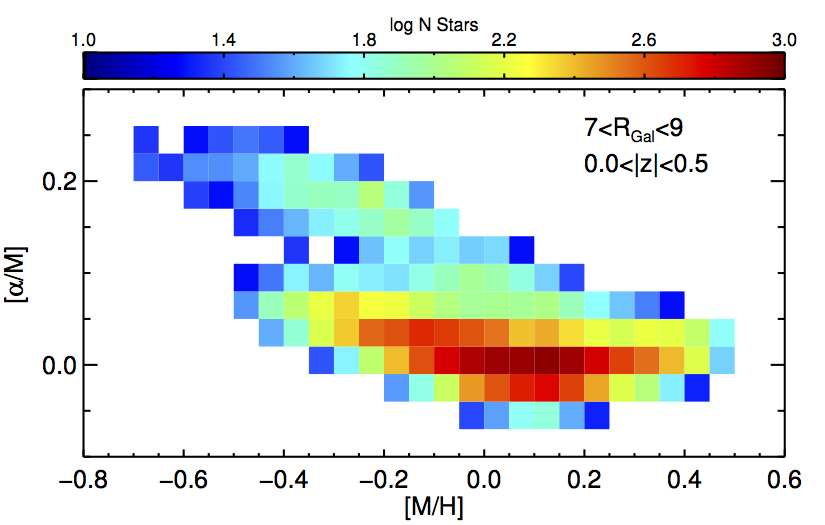}
}
\caption{The [\al/M] vs [M/H] distribution for the $7 <$ \Rgal $ < 9$, $0 < \z < 0.5$ zone colored by the hierarchically modelled mean age (left) and logarithmic number density (right). The sample is binned by 0.05 dex in [M/H] and 0.03 dex in [\al/M]. Only bins that contain at least 15 stars are shown.}
\label{fig:MAA}
\end{figure*}

\subsection{[M/H]-[\al/M]-Age Distribution} \label{sec:maa}

Figure \ref{fig:MAA} combines the age trends presented in Figures \ref{fig:AMR} and \ref{fig:AAR} for the solar neighborhood zone, $7 <$ \Rgal $ < 9$, $0 < \z < 0.5$. In the left panel the mean age is represented by the color of the bins in the [\al/M] vs [M/H] distribution. The right panel is colored by the logarithmic number density of stars. This distribution is binned by 0.05 dex in [M/H] and 0.03 dex in [\al/M]. Bins with fewer than 15 stars are not shown. The typical uncertainty in the mean age in this figure is 0.09 dex, or 0.2 Gyr at 1 Gyr and 1.7 Gyr at 7 Gyr.  Here we provide a qualitative interpretation of this figure in the context of the literature and GCE.

In this figure the youngest stars are concentrated around [M/H] $\sim 0.1$ and [\al/M] $\sim -0.05$, in agreement with the single-element age relations in Figures \ref{fig:AMR} and \ref{fig:AAR}. Note that the lowest [\al/M] stars from Figure \ref{fig:AAR} that had intermediate ages are not present in high enough numbers to appear in Figure \ref{fig:MAA}. Here it is obvious that the high metallicity stars ([M/H] $> 0.3$) have both older ages and higher [\al/M] than the youngest stars. 

The high-\al\, sequence is older and shows little age evolution with [M/H] and [\al/M] along the sequence until solar metallicity, as is seen in Figure \ref{fig:AAR}. This is consistent with the $\sim 1$ Gyr time delay of type Ia supernovae compared the type II, which causes the `knee' in the high-\al\, sequence \citep{Tinsley1979, Matteucci1986}. The high metallicity of the knee, relative to local dwarf galaxies, indicates that the star formation rate was high at that time \citep[see][and references therein]{Tolstoy2009}. The steeper slope of the high-\al\, sequence compared to the low-\al\, sequence is likely caused by a decline in star formation, which results in a dominant contribution of the delayed type Ia supernovae over type II \citep[see][]{Hill2019}. This is consistent with derivations of the local star formation history that find a burst of significant star formation at early time followed by a lull in star formation using thick disc stars \citep[][]{Snaith2014, Haywood2015} and a two-infall model \citep{Spitoni2019}. Most GCE models find the Milky Way disc star formation history can be approximated by a peak in star formation at around 9-10 Gyr ago, followed by a exponential decline in star formation \citep[e.g.][]{Kubryk2015, Rybizki2017, Cote2017a}, also consistent with our results. 

The high [M/H] stars appear to be an extension of the high-\al\, sequence rather than part of the low-\al\, sequence, as debated in \citet{Nidever2014}. Again, it is unlikely that these stars formed locally due to their intermediate age and high [M/H]. In \citet{Anders2018}, these stars are referred to as Inner Disk III and Inner Disk IV based on significant separation of the stars in a t-distributed stochastic neighbour embedding (t-SNE) analysis and their cold orbits.
The low-\al\, sequence is younger and shows more age evolution with [M/H] and [\al/M] along the sequence than the high-\al\, sequence, confirming the more extended star formation history of the low-\al\, sequence proposed previously \citep[e.g.][]{vanDokkum2013, Snaith2014, Rybizki2017}.

The metal-poor end of the low-\al\, sequence is approximately the same age as the metal-rich extension of the high-\al\, sequence. If the high-[M/H] stars formed in the inner disc \citep{Anders2018} and the low-\al\, sequence formed throughout the disc after some gas accretion event \citep[as suggested by e.g. ][]{Chiappini1997, Chiappini2001}, then it is likely that the metal-poor, low-\al\, stars formed in the outer disc and the inner disc was not significantly diluted with the accreted material. 
If the accreted gas did indeed reach the inner disc and the metal-rich stars formed post accretion (coeval with the metal-poor, low-\al\, stars), then inner disc metallicities must have reached [M/H] $> +0.5$ before the gas accretion. Very few stars are have been reported in the literature with such high metallicities \citep[see][]{Ness2013, Bensby2017a, Barbuy2018}. 
Assuming the metal-poor, low-\al\, stars formed locally, then the radial metallicity gradient would have been $-0.15$ dex kpc$^{-1}$ or steeper post gas dilution. This is a much steeper gradient than is measured today. If we use the simple equation
\begin{equation}
\Delta R = \Delta \mbox{[M/H] / gradient},
\end{equation}
assuming the present-day metallicity gradient and a difference in [M/H] of 0.9 dex, then the metal-poor, low-\al\, stars currently found in the solar neighborhood must have formed at least 15 kpc farther out than the metal-rich stars. 

However, if the inner disc gas was not strongly diluted by the merger event, but continued to form stars from gas enriched mainly by {\it in situ} stars, then the metallicity gradient would have been enhanced by the gas accretion and the metal-poor, low-\al\, stars could have migrated a shorter distance. \citet{Minchev2018} find that the radial gradient was likely $-0.15$ dex kpc$^{-1}$ at the earliest times and flattened with time, suggesting a gradient of approximately $-0.1$ dex kpc$^{-1}$ around the time in question. Such a gradient implies that the metal-poor, low-\al\, stars formed 10 kpc outwards of the metal-rich stars.

\citet{Spitoni2019} recently presented a chemical evolution model that suggests a two infall model is sufficient to reproduce the [\al/Fe]-[Fe/H]-age distribution of the solar neighborhood using asteroseismic ages. Figure \ref{fig:MAA} is qualitatively in agreement, but the mean age of the metal-rich stars in this work is 5-6 Gyr compared to 8-10 Gyr as predicted by \citet{Spitoni2019}. The significant age difference between the high-\al\, sequence and the metal-rich stars supports the picture of the metal-rich stars having migrated versus having formed locally during the first of two main epochs on star formation.  From our data, the latter would require a very extended initial star formation period before the gas infall event. More sophisticated comparisons with GCE models are beyond the scope of this work, but will be addressed in future work.

Figure \ref{fig:MAA} is in general agreement with similar figures using LAMOST data in \citet{Xiang2017} and \citet{Wu2018}. Although these [\al/M] vs [M/H] distribution are more extended, the age evolution is quite smooth. In \citet{Xiang2017} the high-\al\, sequence transitions from uniformly old into intermediate ages at a lower [M/H] than in Figure \ref{fig:MAA}. \citet{Wu2018} use asteroseismic ages and find the old ages are present until solar [M/H], in agreement with the present work. \citet{Ness2016b} and \citet{Ho2017} use [C/N]-based ages produced by the {\it Cannon}. \citet{Ness2016b} do not find that the high [M/H] stars are older than the solar abundance stars. \citet{Ho2017} do not extend past [M/H] of 0.2, but is consistent with the present results around solar [M/H].


\section{Conclusions} 
\label{sec:conclusion}

By combining the APOGEE spectroscopic survey with the {\it Gaia} DR2 astrometric catalogue, the sample of red giant stars for which isochrone matching ages are possible has been vastly increased. The sample presented here contains over 75,000 stars with 5 < \Rgal\, < 13 kpc and 0 < \z\, < 2 kpc. The hierarchical modelling method of \citet{Feuillet2016} was used to derive age-abundance trends for [M/H] (MAR) and [\al/M] (AAR) as a function of spatial location in the disc of the Milky Way. This allows us to examine the spatial variations in age-abundance relations on a disc-wide scale for the first time using ages that do not rely on chemical abundance tracers.

There is significant variation in the MAR through the Milky Way disc, Figure \ref{fig:AMR}, an encouraging result for the potential diagnostic power of such observations to constrain large-scale galaxy simulations and models of chemical evolution. These observations suggest that radial migration has a non-negligible effect in the disc plane. From the metallicity of the youngest stars at each \Rgal\, zone in the plane of the disc, the present-day metallicity gradient is measured to be $-0.059 \pm 0.010$ dex kpc$^{-1}$, in agreement with measurements from Cepheids and young field subgiants. The radial metallicity gradient is a key constraint to model of disc evolution \citep[see][]{Stanghellini2019}.  

The outer \Rgal\, zones show evidence in the MAR such that young stars are dominant at larger \z\, in the outer disc than in the inner disc. This is shown explicitly in Figure \ref{fig:flare}. The flaring of the outer disc is predicted by models of Milky Way evolution \citep[see][]{Rahimi2014, Minchev2015} and observed in large surveys \citep[e.g.][]{Ness2016b, Xiang2017, Mackereth2017}.

The AAR, Figure \ref{fig:AAR}, also shows evidence of flaring, but is otherwise fairly consistent in all spatial zones. The relative lack of spatial variation in the AAR is also a strong constraint to models of Milky Way disc evolution. 

The age trends seen in both [M/H] and [\al/M] of the solar neighborhood are nicely recovered in Figure \ref{fig:MAA}, which shows the mean age of mono-[M/H], mono-[\al/M] bins. The high-\al\, sequence is uniformly old with little age evolution until just above solar metallicity. The low-\al\, sequence is younger than the high-\al\, sequence at all [M/H] and shows more significant age evolution. This chemo-age distribution is consistent with previous derivations and models of the Milky Way disc star formation history that find a peak in star formation at around 10 Gyr (forming the high-\al\, sequence) followed by a decline or lull in star formation around 8 Gyr and an extended period of moderate star formation \citep[forming the low-\al\, sequence, e.g.][]{Snaith2014, Kubryk2015, Rybizki2017}. The high-[M/H] stars forming the MAR turn over feature are more likely to be an extension of the high-\al\, sequence, perhaps resulting from a continuation of early star formation in the inner disc, rather than an evolution of the low-\al\, sequence. 

We plan to use these observations, as well as age-abundance relations of individual elements, to constrain models of chemical evolution and Galactic simulations. Extended disc coverage is crucial for more meaningful comparisons to galaxy scale models.

\section*{Acknowledgements}

We thank the anonymous referee for useful suggestions. Many thanks to Ted Mackereth and \'Asa Sk\'ulad\'ottir for helpful discussions. 

DKF and KL acknowledge funds from the Alexander von Humboldt Foundation in the framework of the Sofia Kovalevskaja Award endowed by the Federal Ministry of Education and Research. KL also acknowledges funds from the Swedish Research Council (Grant nr. 2015-00415\_3) and Marie Sklodowska Curie Actions (Cofund Project INCA 600398). DAGH acknowledges support from the State Research Agency (AEI) of the 
Ministry of Science, Innovation and Universities (MCIU) and the European Regional
Development Fund (FEDER) under grant AYA2017-88254-P. ARL acknowledges financial support provided in Chile by Comisi\'on Nacional de Investigaci\'on Cient\'ifica y Tecnol\'ogica (CONICYT) through the FONDECYT project 1170476 and by the QUIMAL project 130001

Funding for the Sloan Digital Sky Survey IV has been provided by the Alfred P.
Sloan Foundation, the U.S. Department of Energy Office of Science, and the
Participating Institutions. SDSS-IV acknowledges support and resources from
the Center for High-Performance Computing at the University of Utah. The SDSS
web site is www.sdss.org.

SDSS-IV is managed by the Astrophysical Research Consortium for the
Participating Institutions of the SDSS Collaboration including the Brazilian
Participation Group, the Carnegie Institution for Science, Carnegie Mellon
University, the Chilean Participation Group, the French Participation Group,
Harvard-Smithsonian Center for Astrophysics, Instituto de Astrof\'isica de
Canarias, The Johns Hopkins University, Kavli Institute for the Physics and
Mathematics of the Universe (IPMU) / University of Tokyo, Lawrence Berkeley
National Laboratory, Leibniz Institut f\"ur Astrophysik Potsdam (AIP),
Max-Planck-Institut f\"ur Astronomie (MPIA Heidelberg), Max-Planck-Institut
f\"ur Astrophysik (MPA Garching), Max-Planck-Institut f\"ur Extraterrestrische
Physik (MPE), National Astronomical Observatories of China, New Mexico State
University, New York University, University of Notre Dame, Observat\'ario
Nacional / MCTI, The Ohio State University, Pennsylvania State University,
Shanghai Astronomical Observatory, United Kingdom Participation Group,
Universidad Nacional Aut\'onoma de M\'exico, University of Arizona, University
of Colorado Boulder, University of Oxford, University of Portsmouth, University
of Utah, University of Virginia, University of Washington, University of
Wisconsin, Vanderbilt University, and Yale University. Collaboration Overview
Start Guide Affiliate Institutions Key People in SDSS Collaboration Council
Committee on Inclusiveness Architects Survey Science Teams and Working Groups
Publication Policy How to Cite SDSS External Collaborator Policy

\bibliographystyle{mnras}
\bibliography{additional,all}

\bsp
\label{lastpage}
\end{document}